\documentclass[12pt, a4paper, parskip, DIV=14]{scrartcl}
\usepackage[english]{babel}
\usepackage[utf8]{inputenc}
\usepackage[T1]{fontenc}  \usepackage{textcomp, booktabs, multirow}
\usepackage[hidelinks]{hyperref}
\usepackage{graphicx, grffile}  \usepackage[separate-uncertainty=true]{siunitx}
\usepackage{amsmath, xspace}

\usepackage{lmodern}
\usepackage{doi, authblk}
\usepackage[round]{natbib}
\usepackage[format=plain, indention=0.5cm, font=small]{caption}

\usepackage{xcolor}

\catcode`\^^M=10

\newcommand{\ly}{\ensuremath{\mathit{ly}}\xspace}
\newcommand{\lobs}{\ensuremath{\ly\ind{obs}}\xspace}
\newcommand{\ind}[1]{_\text{#1}}
\newcommand{\parent}[1]{\ensuremath{\left(#1\right)}\xspace}

\newcommand{\fof}{\!\parent}

\newcommand{\Le}{\ensuremath{L\ind e}\xspace}
\newcommand{\Lez}{L_{\text e0}}
\newcommand{\Lm}{\ensuremath{L\ind m}\xspace}
\newcommand{\Lmz}{L_{\text m0}}
\newcommand{\Ls}{\ensuremath{L\ind s}\xspace}
\newcommand{\ts}{t\ind s}
\newcommand{\tsz}{t_{\text s0}}
\newcommand{\tl}{t\ind l}
\newcommand{\Tl}{T\ind l}
\newcommand{\f}[2]{f_{#1#2}}
\newcommand{\Ie}{I\ind e}
\newcommand{\Iez}{I_{\text e0}}
\newcommand*\diff{\mathop{}\!\mathrm{d}}

\hypersetup{pdfauthor={Tom Eulenfeld and Christoph Heubeck}, pdftitle={Moon's orbit 3.2 billion years ago}}

\title{\LARGE Constraints on Moon's orbit 3.2 billion years ago from tidal bundle data}
\author{Tom Eulenfeld}
\author{Christoph Heubeck}
\affil{\small Friedrich Schiller University Jena, Institute for Geosciences, Burgweg 11, 07749 Jena, Germany\newline contact: tom.eulenfeld@uni-jena.de}
\date{{\small December 20, 2022}}
\begin{document}
\maketitle
\vspace{-1.7cm}
\begin{center}
\color{gray} \footnotesize An edited version of this article was published by\\\emph{Journal of Geophysical Research: Planets}, doi:~\href{http://doi.org/10.1029/2022JE007466
}{10.1029/2022JE007466}.
\end{center}
\vspace{-0.5cm}

\begin{abstract}\noindent
The angular momentum of the Earth-Moon system was initially dominated by Earth's rotation with a short solar day of around 5 hours duration.
Since then, Earth gradually transferred angular momentum through tidal friction to the orbit of the Moon, resulting in an increasing orbital radius and a deceleration of Earth's rotation. Geologic observations of tidal deposits can be used to verify and constrain models of lunar orbital evolution.

In this work we reexamine the oldest tidal record suitable for analysis
from the Moodies Group, South Africa, with an age of 3.22 billion years.
Time frequency analysis of the series of thicknesses of the sandstone-shale layers yields a periodicity of 15.0 layers, taking into account the possibility of missing laminae.

Assuming a mixed tidal system, the duration of two neap-spring-neap cycles was 30.0 lunar days for dominant semidiurnal or 30.0 sidereal days for dominant diurnal tides.

We derive the relationship between this observation and the past Earth-Moon distance and re-visit related published work.

We find that the Earth-Moon distance 3.2 billion years ago was about 70\% of today's value. The Archean solar day was around 13 hours long.
The ratio of solar to lunar tide-raising torque controls the leakage of angular momentum from the Earth-Moon system, but deviation from the assumed ratio of 0.211 results in only moderate changes.

A~duration of a postulated 21-hour atmospheric resonance shorter than 200 million years would be consistent with our observation; it would significantly alter Earth-Moon distance.

\newpage
\paragraph{Plain language summary}
After its formation 4.5 billion years ago, the Moon circled Earth in a low orbit  while Earth rotated faster than today around its axis.
In the course of time, the Moon gradually evolved to a higher orbit while the rotation of Earth slowed due to the frictional effect of tides. Theoretical models can describe the evolution of the distance between Earth and the Moon with time until today.
Counting the thickness of thin sandstone-shale couplets of known age, which are layered due to tides, can constrain these models.
In this work we reexamine the oldest of these geological records in the Moodies Group of South Africa, with an age of 3.2 billion years. The thickness of layers changes with a periodicity of 15 layers which is assumed to originate from varying strengths of currents between successive spring tides.
Kepler's third law and the law of conservation of angular momentum allow us to derive the parameters of the lunar orbit from this measurement. According to our analysis, the Earth-Moon distance was around 70\% of today's value 3.2 billion years ago. The faster rotation rate of Earth resulted in a length of day of around 13 hours.
\newline\newline
Key points:
\begin{itemize}
\item Time frequency analysis yields 30.0 layers per two neap-spring-neap cycles, taking missing laminae in the tidal record into account
\item Earth-Moon distance of ca.\ 70\% of today's value 3.2 billion years ago results in a solar day of 13 hours duration
\item Duration of 21-hour atmospheric resonance for \textless 200 million years is consistent with our observation, alters estimate of Earth-Moon distance
\end{itemize}
\vspace{1ex}
\end{abstract}

\section{Introduction}
\label{sec:intro}
The impact of a Mars-sized body on Earth 4.5 billion years ago is the most accepted theory for the origin of the debris that formed the Moon \citep[and many others]{Hartmann1975, Canup2001}.
The angular momentum of the Earth-Moon system was initially dominated by Earth's rotation which resulted in a short solar day of around 5~hours \citep[e.g.][]{Goldreich1966}. Since then, Earth gradually transferred angular momentum through tidal friction to the orbital angular momentum of the Moon; to a lesser extent angular momentum was transferred to the Sun system (and is considered ``lost'' for the Earth-Moon system). After the formation of liquid oceans on Earth, at the latest around 4.2~billion years ago, tidal friction of the oceans also contributed to this effect. As result, the Earth's day lengthened while Moon gradually moved away from the Earth, increasing its orbital period. Moon's present recession rate of \SI{3.8}{cm} per year, determined by Apollo's Lunar Laser Ranging \citep{Dickey1994, Williams2016}, is a consequence of this ongoing process.\par

Key parameters describing the orbital mechanics of the Earth-Moon system include the distance between Moon and Earth, the rotation period of Moon around Earth and the rotation speed of Earth around its own axis. The relationship between these parameters can be derived by Kepler's third law and the conservation of angular momentum in a system including Moon, Earth and Sun.
Determining the temporal evolution of these parameters is more difficult, because the extent of tidal friction is related to several parameters, most importantly the distance between Earth and Moon, Earth's rotation rate, and the changing size and configuration of land masses and oceans \citep[e.g.][]{Nance2014}.
Several authors traced tidal dissipation with time in order to infer the temporal evolution of the lunar orbit \citep[e.g.][]{Webb1982, Daher2021}. These models can be constrained by geologic observations of tidal laminae with known depositional ages. Tidal laminae are generated by the alternating deposition of sand and mud during phases of strong and weak currents, respectively, at different times of the tidal cycle.
Determining their rhythmicities works best for long records of tidal deposits, ideally spanning several years, because thus several independent orbital parameters can be inferred and verified against each other. For example, long tidal records of Precambrian glaciogenic strata with an age of ca.\ 650 million years in South Australia permit to determine independently the number of synodic months per year, the number of lunar days (defined as the duration between two successive culminations of the Moon) per synodic month and the length of the lunar nodal cycle in years; all three parameters leading to the same Earth-Moon distance of approximately 97\% of today's distance \citep{Williams1989, Deubner1990}. Data measured in tidal deposits of far older strata are of great interest because they could constrain the poorly known early lunar orbit.\par

\citet{Eriksson2000} contributed the earliest data point in this collection by analyzing the rhythmicity of tidal bundles from an exposed outcrop in the Moodies Group of the Barberton Greenstone Belt, South Africa, about 3.22 billion years old. A tidal bundle is a single thin sandstone-shale combination preserved in inclined foreset strata of large fossilized subtidal sand dunes. The varying thicknesses of these foreset strata is thought to be due to the variable speed and duration of tidal currents.  Because shale laminae at this site are very thin in comparison to the sandstone beds, we will in the following refer to a bundle as a single layer. 
By frequency analysis, \citet{Eriksson2000} originally calculated a periodicity of 13.1 layers but recalculated a periodicity of 9.3 after filtering and removing an assumed record of subordinate semidiurnal tides, a processing step disapproved by \citet{Heubeck2022} as unwarranted.

Based on the periodicity of 9.3, \citet{Eriksson2000} concluded that the ``anomalistic month at \SI{3.2}{Ga} was closer to 20 days than the present 27.55 days''. Because semidiurnal tides are related to the lunar day and two neap-spring-neap cycles correspond to a synodic month for semidiurnal tides, \citet{Eriksson2000} would have been better served by reporting this result as the number of lunar days in a synodic month. \citet{Heubeck2022} pointed out that a synodic month of only 20 lunar days was unrealistically short and did not conform to accepted models of lunar orbital evolution.
Applying the equations which will be presented in this study, the Earth-Moon distance would only have been 42\% of today's value.
Because the tidal dissipation is proportional to the inverse of the sixth power of the Earth-Moon distance \citep[e.g.][]{Goldreich1966b}, this orbit could only be occupied by a newly formed Moon. A value of 26.2 lunar days per synodic month, derived from the unfiltered data set of \citet{Eriksson2000}, appears to be more realistic 3.2 billion years ago. \citet{Heubeck2022} derived a value of 28 lunar days per synodic month from data measured at the same outcrop and interpreted this value as a lower bound due to the possibility of missing laminae.
\par

\Citet{Azarevich2017} reported that \citet{Eriksson2000} calculated an Earth-Moon distance of 75\% to 80\% of today’s value 3.2 billion years ago. Such a statement, however, cannot be found in \citet{Eriksson2000} so that it remains unclear how \citet{Azarevich2017} obtained these values. They correspond poorly to results we obtain by applying relevant equations of our study to the original observations of \citet{Eriksson2000}.
\par

The periodicity in the Moodies tidal record can only be quantified in the spectral domain because of the high variability in plots of laminae thicknesses.
Furthermore, the record is special because in previous analyses the number of layers per neap-spring-neap cycle corresponded directly to the number of lunar days in a synodic month \citep[e.g.][]{Visser1980, Archer1996}. This is consistent with a semidiurnal tidal system resulting in two periods of weak and strong currents per lunar day and thus in two deposited sandstone-shale bundles per lunar day (a 1:2 correspondence). In contrast, a 1:1 correspondence applied to the Moodies tidal record would have resulted in a synodic month consisting of only 14 lunar days \citep[or of 10 lunar days of][respectively]{Eriksson2000} which is again too short for realistic models of lunar orbital evolution. A 1:2 correspondence, as applied in \citet{Eriksson2000} and \citet{Heubeck2022}, could be explained if the deposition was due to a mixed tidal system possibly with dominating diurnal contribution. However, the interpretation of the observed periodicity would be different.\par

The objective of this article is three-fold.
In section~\ref{sec:tides}, we discuss consequences of interpreting geological tidal records as semidiurnal and diurnal in origin, respectively \citep[see also][]{Kvale2006}.
In section~\ref{sec:specs}, we perform a time-frequency analysis on the best data set from the Moodies outcrop in question \citep{Heubeck2022}, taking also in account the possibility of missing bundles.
In section~\ref{sec:orbit}, we derive orbital parameters solely from the number of layers per two neap-spring-neap cycles \citep{Runcorn1979}, consider a tidal system with dominating diurnal tide and add a correction for the change in Earth's moment of inertia.

In section~\ref{sec:final}, constraints on the lunar orbit 3.2 billion years ago are summarized and underlying assumptions and implications for models of lunar orbital evolution are discussed.

\section{Implications of semidiurnal versus diurnal tides for the time-series interpretation of tidal deposits}
\label{sec:tides}

\begin{figure}
\centering
\includegraphics[width=0.9\textwidth]{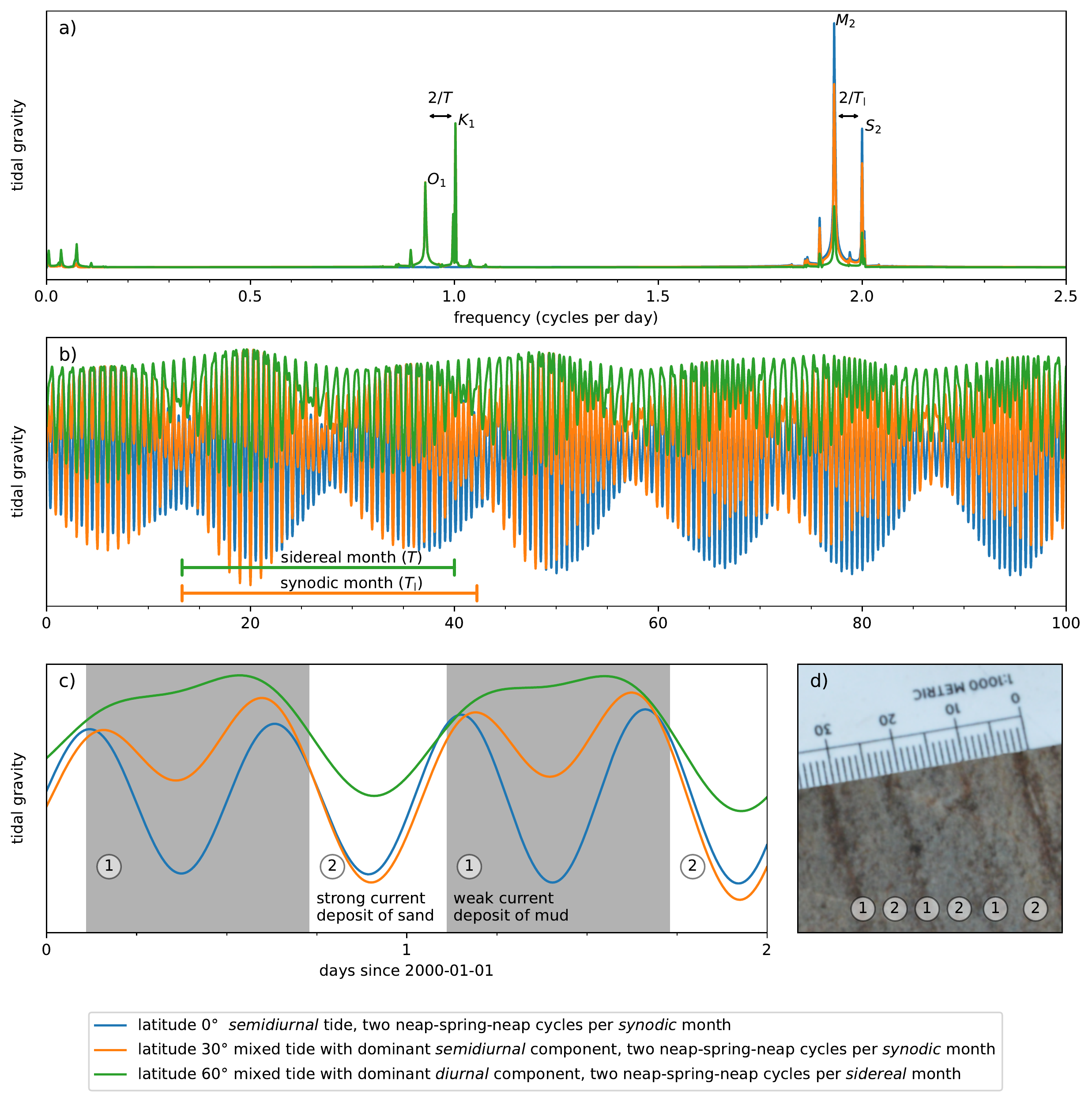}
\caption{
Spectrum of tidal gravity (a) and time series of tidal gravity for 100 days (b) calculated at three latitudes (longitude 0°) for present Earth.
Depending on latitude the tidal gravity illustrates a purely semidiurnal tide (blue), a mixed tide with dominating semidiurnal component (orange) or a mixed tide with dominating diurnal component (green).
The neap-spring-neap cycles are clearly recognizable in b) and their durations are indicated by horizontal lines; they are induced by the beat between neighboring semidiurnal tidal constituents M$_2$ and S$_2$ or neighboring diurnal constituents O$_1$ and K$_1$. The beat frequency is given by the frequency difference between the corresponding tidal constituents.
Therefore the neap-spring-neap cycle is of different duration for semidiurnal and for diurnal tides.\newline
c) Cutout of the time series in b) displaying the first two days. The deposition of a single sand-mud layer per day is interpreted as being due to a mixed tide with dominant diurnal or semidiurnal component, that is with one period of strong and one period of weak currents each day. Different flow channels for rising and falling tide could also be possible. In this case sand is only deposited in periods of either rising or decreasing gravity.\newline
d) Outcrop photograph of the Moodies tidal record showing alternating layers of shale and sandstone, corresponding to phases of the tidal currents (encircled numbers in c) and d)). Layers are measured from the middle of a shale layer to the middle of the subsequent shale layer \citep{Heubeck2022}. The scale is copied from another part of the same photograph.
}
\label{fig:tides}
\end{figure}

In this section, we address the question in which cases a tidal system may result in a single bundle (in the following: a layer) per day, resulting in a 1:2 correspondence as applied in \citet{Eriksson2000} and \citet{Heubeck2022} for the Moodies tidal record.
We therefore introduce the variable $\ly$, defined as the \emph{number of days per two neap-spring-neap cycles}. ``Day'' is not clearly defined here and refers to the period of the tidal system at work. Loosely speaking, $\ly$ corresponds to the number of days in a month. Further we introduce the observable $\lobs$ which is defined as the \emph{observed number of layers per two neap-spring-neap cycles}. The ratio of $\lobs$ to $\ly$ is determined by the tidal system. As mentioned in most earlier publications, $\lobs/\ly=2$.\footnote{The 1:1 correspondence translates to a ratio $\lobs/\ly=2$, because the \ly observables are defined relative to \emph{two} neap-spring-neap cycles.} In the following we discuss under which circumstances a ratio $\lobs/\ly=1$ could occur in the Moodies tidal record and we want to clarify the ambiguous term ``days per month'' and its relationship to the observable $\lobs$.\par

Figure~\ref{fig:tides} shows three synthetic time series of the tidal gravity for today's solid Earth together with their spectra at different latitudes, corresponding to a purely semidiurnal system (latitude 0°), a mixed tidal system with dominating semidiurnal tide (latitude 30°) and a mixed tidal system with dominating diurnal tide (latitude 60°). The synthetic tidal signals were generated with ETERNA predict via its Python wrapper pygtide \citep{Wenzel1996, Rau2022}. Ocean tides are also influenced by several other factors, for example by ocean resonances and the shape of the shore, but in any case we can discuss the characteristics of tidal systems with dominating semidiurnal and  diurnal component using the resulting time series.
The main contribution to semidiurnal tides are due to the principal semidiurnal tidal constituents of Moon and Sun ($M_2$ and $S_2$, respectively). The corresponding 
neap-spring-neap cycle is completed twice per synodic month $\Tl$ and arises from the beat between the two semidiurnal constituents. Its frequency is given by the difference of the frequencies of the semidiurnal constituents that are in turn related to the length of the lunar day $\tl$ and of the solar day $\ts$ \citep[e.g.][]{Doodson1921}

\begin{eqnarray}
\f M2 = 2/\tl \,, &  \f S2 = 2/\ts \,, & \f S2 - \f M2 = 2 /\Tl \,.
\end{eqnarray}

Because the lunar $M_2$ constituent generally has a higher amplitude than the solar $S_2$ constituent, individual tide waves have a period of half a lunar day $\tl/2$, corresponding to the inverse of the frequency $\f M2$. $\ly$ for semidiurnal tides or for mixed tides with dominant semidiurnal contribution is therefore given by the number of lunar days per synodic month

\begin{equation}
\ly_2 = \Tl/\tl \,. \label{eq:ly2}
\end{equation}

\citet{Visser1980} and others observed two tidal bundles per lunar day ($\lobs/\ly=2$) and interpreted this ratio to be due to the inequality of the twice-daily tidal currents or due to tidal currents occupying different flow channels.

This explanation corresponds to the blue line in figure~\ref{fig:tides}c and the consideration of only increasing or decreasing periods of gravity as periods of strong current resulting in a deposition of sand.

\citet{Eriksson2000} and \citet{Heubeck2022}, in contrast, both interpreted each tidal bundle to represent one lunar day ($\ly/\lobs=1$), reflecting a mixed tidal system with a dominant semidiurnal component and only one period of strong and weak currents per day (orange line in figure~\ref{fig:tides}c).

The ratio $\ly/\lobs=1$ can also be explained as representing a mixed tidal system with dominant diurnal component (green line in figure~\ref{fig:tides}c). A similar argument could be held up as in \citet{Visser1980}. The representation of one day by a single period of strong and a single period of weak currents (figure~\ref{fig:tides}c) takes into account that the deposition of a mud layer separating the sand layers needs a certain minimal duration of slack current. Summing up, deposition of a single layer per day as interpreted for the Moodies outcrop can be best explained by a tidal system with a dominant diurnal component, but a mixed tidal system with dominant semidiurnal component cannot be excluded. Figures~\ref{fig:tides}c and d illustrate this argument. A diurnal tidal system may indicate a high latitude of the location at the time of deposition.\par

It is important to understand the meaning of $ly$ for diurnal tides \citep[e.g.][]{Kvale2006}. The largest diurnal constituents are $K_1$ and $O_1$ (figure~\ref{fig:tides}a). Their frequencies and the difference of their frequencies related to the diurnal neap-spring-neap cycle are given by \citep[e.g.][]{Doodson1921}

\begin{eqnarray}
\f K1 = 1/t \,, &  \f O1 = 1/t - 2/T \,, & \f K1 - \f O1 = 2 / T
\end{eqnarray}

with sidereal day $t$ and sidereal month $T$.\footnote{Actually the difference of frequencies of $O_1$ and $K_1$ is related to the tropical month which takes into account the precession of Earth's rotation axis with a period of approximately 26000~years. Tropical month and sidereal month differ by less than 10~seconds and are considered the same in this study. Similarly, the difference between Earth's rotation rate with respect to the fixed stars and relative to the precessing equinox is negligible; the period of Earth's rotation is considered a sidereal day in both cases.}
The $K_1$ constituent incorporates effects of Sun and Moon and therefore has a larger amplitude than the purely lunar constituent $O_1$. The daily tidal deposit is therefore related to the sidereal day and $\ly$ of diurnal tides is given by the number of sidereal days per sidereal month

\begin{equation}
\ly_1 = T/t \,. \label{eq:ly1}
\end{equation}

In summary, layers in the Moodies tidal record are interpreted as day deposits ($\lobs/\ly=1$). The interpretation of the value \ly (``number of days in a month'') depends on the dominant tide: It may represent the number of lunar days per synodic month in case of semidiurnal tides (equation~\ref{eq:ly2}) or the number of sidereal days per sidereal month in case of diurnal tides (equation~\ref{eq:ly1}).

\section{Time-frequency analysis of the Moodies tidal record}
\label{sec:specs}

\begin{figure}
\centering
\includegraphics[width=0.9\textwidth]{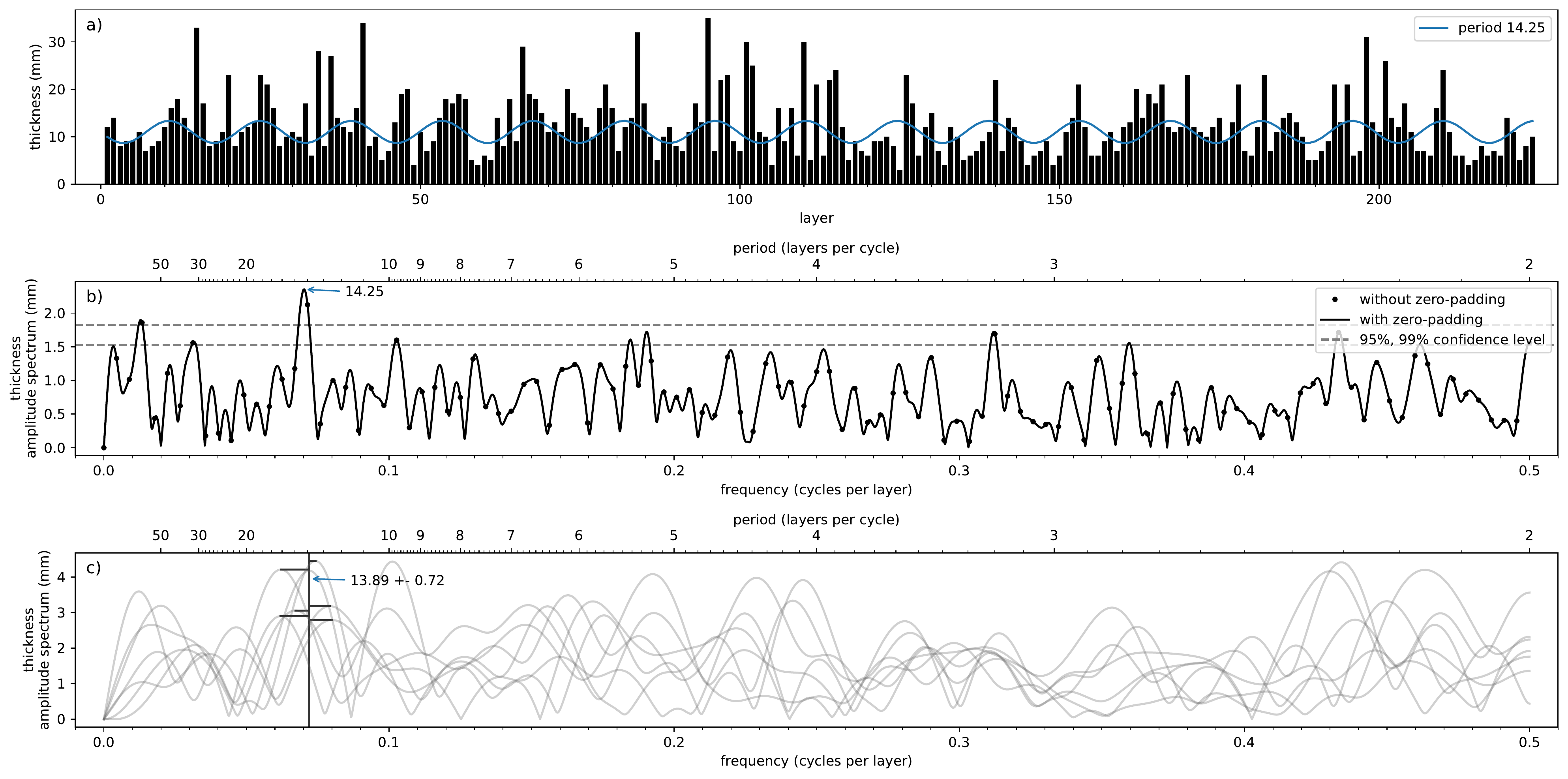}
\caption{
Figure adapted from figure~15 of \citet{Heubeck2022}.\newline
a) Composite data set with 224 measurements described in \citet{Heubeck2022}.\newline
b) Amplitude spectrum of the original data set (black points) and the zero-padded data set (continuous line). The frequency bin width and frequency resolution is 0.0045 cycles per layer for the Fourier transform of the original data set. The frequency sampling is greatly enhanced by zero-padding, contrary to the frequency resolution, which is the same as in the unpadded spectrum. The 95\% and 99\% confidence levels for which the power spectrum is different from white noise is indicated.
The harmonic corresponding to the peak with a period of 14.25 layers per cycle is displayed in panel a).\newline
c) Amplitude spectra of subsets of the data of length 50 with an overlap of 25 data points (gray lines) with zero-padding applied. The black vertical line displays the median of the period estimates of the peak of interest for the individual spectra. Residuals are displayed with black horizontal lines. The frequency resolution is mediocre at 0.02 cycles per layer. 
}
\label{fig:spec}
\end{figure}

\begin{figure}
\centering
\includegraphics[width=\textwidth]{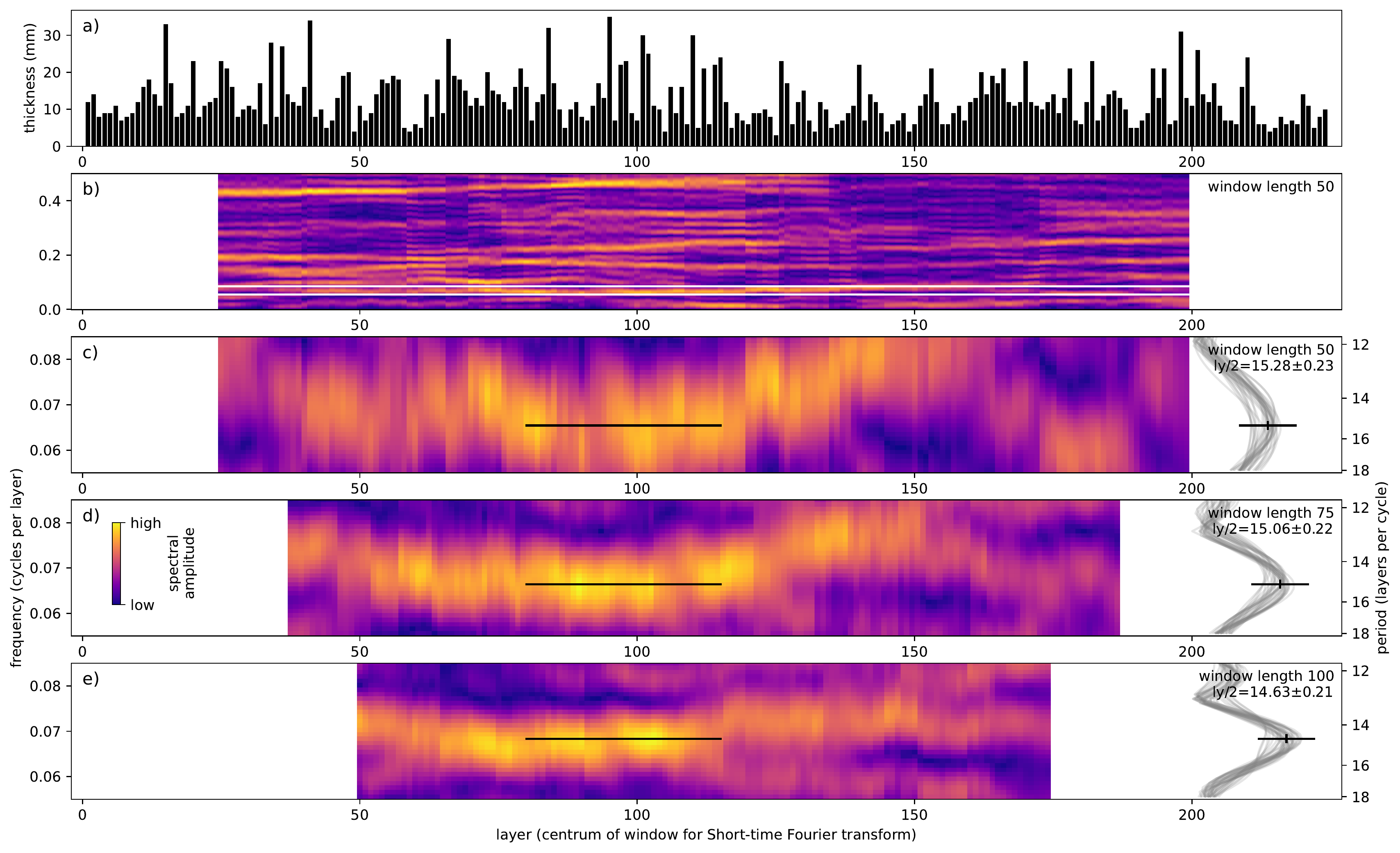}
\caption{
Time-frequency analysis of the Moodies outcrop.\newline
a) Composite data set with 224 measurements, see figure~\ref{fig:spec}a.\newline
b) Spectrogram of the data set displayed in a). The Short-time Fourier transform (STFT) is calculated for a window length of 50 layers with a sliding of 1 layer.\newline
c) Zoom to the relevant frequency range 0.55 to 0.85 cycles per layer (white lines) of the spectrogram in b). The Fourier transforms corresponding to the layers 80 to 115 show the strongest amplitude at the 0.065 cycles per layer peak. For these parts of the tidal record the frequency of the peak is also slightly shifted to lower values. In the right part of the figure, we display the 36 individual Fourier transforms for this part of the signal. Black lines denote the median of the maxima of these 36 spectra and the median absolute deviation (MAD) of the maxima around their median. The median and MAD in terms of period (layers per cycle) are shown in the upper right corner of the panel.\newline
d) Same plot as in c) with a window length of 75 samples in the STFT.\newline
e) Same plot as in c) with a window length of 100 samples in the STFT.
}
\label{fig:spectrograms}
\end{figure}

Figure~\ref{fig:spec}, adapted from \citet{Heubeck2022}, shows the time series and spectrum of the data set from the tidal deposit described by \citet{Eriksson2000} and reexamined by \citet{Heubeck2022}.
The composite data set of layer thickness is displayed in figure~\ref{fig:spec}a and the corresponding amplitude spectrum in figure~\ref{fig:spec}b. The largest peak observed at a period of 14.2 layers per cycle does not arise from white noise with a confidence level higher than 99\%. \citet{Heubeck2022} argue that this spectral peak represents the number of layers deposited in a neap-spring-neap cycle.

The periodicity of 14.2 is not constant throughout the full data set but varies within subsets (figure \ref{fig:spec}c) suggesting that layers may be missing due to non-deposition or erosion. 

This observation lies at the base of the time-frequency analysis shown in figure~\ref{fig:spectrograms}. The same fluctuation of the 14.2 peak, observable in Fourier transforms of subsets of the data, is also observable in the spectrograms of figure ~\ref{fig:spectrograms}c-e.

We select a segment between layer 80 and layer 115 in which the frequency peak at 0.065 cycles per layer is strongest and shifted to lower frequencies. These data points are used as midpoints for individual Short-time Fourier transforms of length 50, 75 and 100 data points, respectively (figure~\ref{fig:spectrograms}b-d, right). Medians of the position of the relevant frequency peak lie at periods 15.3, 15.1 and 14.6 layers per cycle for a window length of 50, 75 and 100 data points, respectively. The average of these three individual values, 15.0 layers per neap-spring-neap cycle, will be used below. The value 15.3 of the Short-time Fourier transform, derived from the window length of 50, and the value of 14.2, taken from \citet{Heubeck2022}, will be used as upper and lower bound to this mean, respectively. Consequently, the observed number of layers per two neap-spring-neap cycles will be $\lobs = 30.0$ with lower and upper bounds of 28.4 and 30.6.

\section{Moon's orbit from short data sets of tidal laminae}
\label{sec:orbit}

\begin{table}
  \caption{Used variable names and constants.}
  \label{tab:var}
  \centering
  \begin{tabular}{ll}\toprule
    $\ly$ & number of ``days'' per two neap-spring-neap cycles\\
    $\lobs$ & number of layers per two neap-spring-neap cycles\\
    $r$ & Moon's semi-major axis, mean Earth-Moon distance, orbital radius\\
    $a$ & ratio of Moon's semi-major axis to its current value\\
    $t$ & period of Earth's rotation or duration of sidereal day\\
    $t\ind l$ & duration of lunar day\\
    $t\ind s$ & duration of solar day\\

    $T$ & lunar orbital period or duration of sidereal month\\
    $T\ind l$ & duration of synodic month\\

    $\Le$ & angular momentum of Earth's rotation\\
    $\Lm$ & angular momentum of Moon's orbit\\
    $\Ls$ & angular momentum of Sun-Earth system\\

    $m$ & Moon's mass\\

    $\Ie$, $\Iez$ & Earth's polar moment of inertia, today's value\\
    $\ly_2=T\ind l / t\ind l$ & $\ly$ for semidiurnal tides, number of lunar days in synodic month\\
    $\ly_1=T / t$ & $\ly$ for diurnal tides, number of sidereal days in sidereal month\\
    $\alpha_0=\Lez/\Lmz=0.203$ & today's ratio of angular moments of Earth and Moon\\
    $\beta_0=0.211$ & today's ratio of solar tide-raising torque and lunar tide-raising torque\\
    $t_0 = \SI{86164}s$ & today's duration of sidereal day\\        
    $\tsz = \SI{86400}s$ & today's duration of solar day\\    
    $T_0 = 27.32\tsz$ & today's duration of sidereal month\\
    $Y=365.24\tsz$ & duration of a year (assumed constant)\\
    $r\ind e$ & Earth's radius\\
    $G$ & gravitational constant\\      
    $k_2^{\text f}=0.93$ & Earth's fluid Love number around the rotation axis\\
    $\gamma=0.00215$ & factor accounting for the change in $\Ie$ due to different rotation speed\\
    $\mu=1$ &  relative change in $\Ie$ independent from the rotational effect\\
    
    \bottomrule
  \end{tabular}
\end{table}

In this section, we assume that the value of \ly (``number of days in a month'') was successfully determined. For diurnal tides $\ly= \ly_1$ is the number of sidereal days in the sidereal month, and for semidiurnal tides $\ly = \ly_2$ is the number of lunar days in the synodic month, as discussed in section~\ref{sec:tides}.

From this single observable we derive the lunar orbit, i.e. the Earth-Moon distance, the lunar orbital period and the duration of a solar day. We closely follow the derivation of \citet{Runcorn1979} allowing for a lack of angular momentum from the Earth-Moon system to the Sun-Earth system.
Like \citet{Runcorn1979} we disregard the eccentricity of the lunar orbit (current mean 0.055) as well as its inclination towards Earth's orbit around the Sun (current mean 5.1°). The tilt of Earth's rotation axis, which is currently 23.4°, is also neglected.
Differences to a similar derivation presented in \citet{Coughenour2013} are identified.
All symbols used in this derivation are listed in table~\ref{tab:var}, but are also introduced in the text.\par

Kepler's third law is given by

\begin{equation}
T = T_0 a^{3/2}
\label{eq:Kepler}
\end{equation}

with lunar orbital period or sidereal month $T$, its present value $T_0$ and $a=r/r_0$ being the ratio of Moon's semi-major axis $r$ (the mean Earth-Moon distance, orbital radius) to today's value $r_0$ ($\approx \SI{384000}{km}$). The angular moments of Earth and Moon relative to current values are given by

\begin{align}
\Le &= \frac{2\pi \Ie}t = \Lez\frac{t_0}t\frac\Ie\Iez\label{eq:Le}\\
\Lm &= \frac{2\pi m r^2}T = \Lmz a^{1/2} \label{eq:Lm}

\end{align}

with Earth's polar moment of inertia $\Ie$, its current value $\Iez$, length of sidereal day $t$, its current value $t_0$ and Moon's mass $m$.

Due to friction, the tidal bulges caused by the gravitation of the Moon do not align with the Earth-Moon axis \citep[e.g.][]{Goldreich1966b}. The gravitational torque the Moon exerts on these bulges is responsible for part of the deceleration of Earth's rotation; the gravitational torque exerted by these bulges on the Moon is responsible for pushing the Moon to a higher orbit. Similarly, the tidal bulges caused by the gravitation of the Sun do not align with the Sun-Earth axis. The gravitational torque the Sun exerts on the tidal bulges is responsible for the remaining deceleration of Earth's rotation and for pushing Earth to a higher orbit around the Sun. The second effect is negligible in this calculation and the length of the year is assumed to be constant. Because of the conservation of angular momentum, changes in angular momentum of the Earth-Moon system $\Le+\Lm$ and in the angular momentum of the Sun-Earth system \Ls cancel each other:

\begin{equation}
\diff\Le + \diff\Lm + \diff\Ls = 0 \label{eq:consttotal}
\end{equation}

Angular momentum from Earth's rotation \Le is transferred to the angular momentum of Moon's orbit \Lm and to the angular momentum of the Sun-Earth system \Ls. The ratio of these two momentum transfers $\diff\Ls / \diff\Lm$ equals the ratio of torques exerted by Sun and Moon on the tidal bulges. Assuming the same lag angles for the tidal bulges induced by Sun and Moon, the ratio of torques equals the square of the ratio of Sun's and Moon's mass multiplied by the sixth power of the ratio of the distances Earth-Moon and Earth-Sun \citep[e.g.][]{Goldreich1966b}; $\beta_0=0.211$ is today's ratio between the solar tide-raising torque and the lunar tide-raising torque. Because the change in the Earth-Sun distance due to momentum transfer to Earth's orbit is negligible, the ratio between the solar tide-raising torque and the lunar tide-raising torque for different lunar orbits is given by $\beta_0 a^6$ and thus:

\begin{equation}
\diff\Ls = \beta_0 a^6 \diff\Lm \label{eq:constLs}
\end{equation}

The validity of the assumptions underlying equation~(\ref{eq:constLs}) will be discussed later.
Combining equation (\ref{eq:consttotal}) and (\ref{eq:constLs}) gives

\begin{equation}
\diff\Le = -\diff\Lm \parent{1 + \beta_0 a^6}\,. \label{eq:cons1}
\end{equation}

By dividing by $\Lmz$, substituting $a = a'$ and integrating over $a'$ with
$\diff\Lm = \Lmz a'^{-0.5}\diff a'/ 2$ it can be derived that

\newcommand{\exprX}{1 + \frac 1{13}\beta_0 - a^{0.5} - \frac 1{13} \beta_0 a^{6.5}}
\newcommand{\X}{X\fof{a, \beta_0}}

\begin{align}
\alpha_0\parent{\frac{t_0}t\frac\Ie\Iez - 1} &=  \X \text{ with} \label{eq:integral}\\
\X &=  - \int_{a_0}^a \frac 12 a'^{0.5} + \frac 12 \beta_0 a'^{5.5} \diff a' \text{ and}\nonumber\\
\X &=  a_0^{0.5} + \frac 1{13}\beta_0a_0^{6.5} - a^{0.5} - \frac 1{13} \beta_0 a^{6.5}\label{eq:X}\\
   &=  1 + \frac 1{13}\beta_0 - a^{0.5} - \frac 1{13} \beta_0 a^{6.5}\nonumber
\end{align}

with today's ratio of angular moments of Earth and Moon $\alpha_0=\Lez/\Lmz=0.203$,

and finally

\begin{equation}
\frac{\alpha_0 t_0}t\frac\Ie\Iez = \alpha_0 + \X \label{eq:cons2} \,.
\end{equation}

We derived this equation \citep[also presented in][and others]{Runcorn1979, Deubner1990} to bring out the assumptions and to apply it to diurnal tidal records. Equation~(\ref{eq:cons2}) is an equation of two unknowns $t$ and $a$ assuming a constant moment of inertia $\Ie=\Iez$. Finally, we need to express $t$ by the known variable $\ly$. For diurnal tides this is straight-forward from equation~(\ref{eq:ly1}):

\begin{equation}
\frac 1t = \frac{\ly_1}T \label{eq:t1}
\end{equation}

For semidiurnal tides using equation (\ref{eq:ly2}) follows:

\begin{equation}
\frac 1t = \frac 1T + \frac 1\tl = \frac 1T + \frac{\ly_2}\Tl = \frac 1T + \ly_2\parent{\frac 1T - \frac 1Y} \label{eq:t2}
\end{equation}

Equation (\ref{eq:Kepler}) is inserted into equation (\ref{eq:t1}) and (\ref{eq:t2}), respectively. Equation (\ref{eq:t1}) respective (\ref{eq:t2}) can be inserted into equation (\ref{eq:cons2}) and and solved for $ly_1$ respective $ly_2$ as a function of $a$:

\newcommand{\exprT}{T_0a^{3/2}}
\newcommand{\expra}{\frac\X{\alpha_0} + 1}

\begin{align}
\ly_1 &= \parent\expra \frac\exprT{t_0}\frac\Iez\Ie \label{eq:ly1a}\\
\ly_2 &= \parent{\parent\expra \frac 1{t_0}\frac\Iez\Ie - \frac 1\exprT} / \parent{\frac 1\exprT - \frac 1Y} \label{eq:ly2a}
\end{align}

\begin{figure}
\centering
\includegraphics[width=\textwidth]{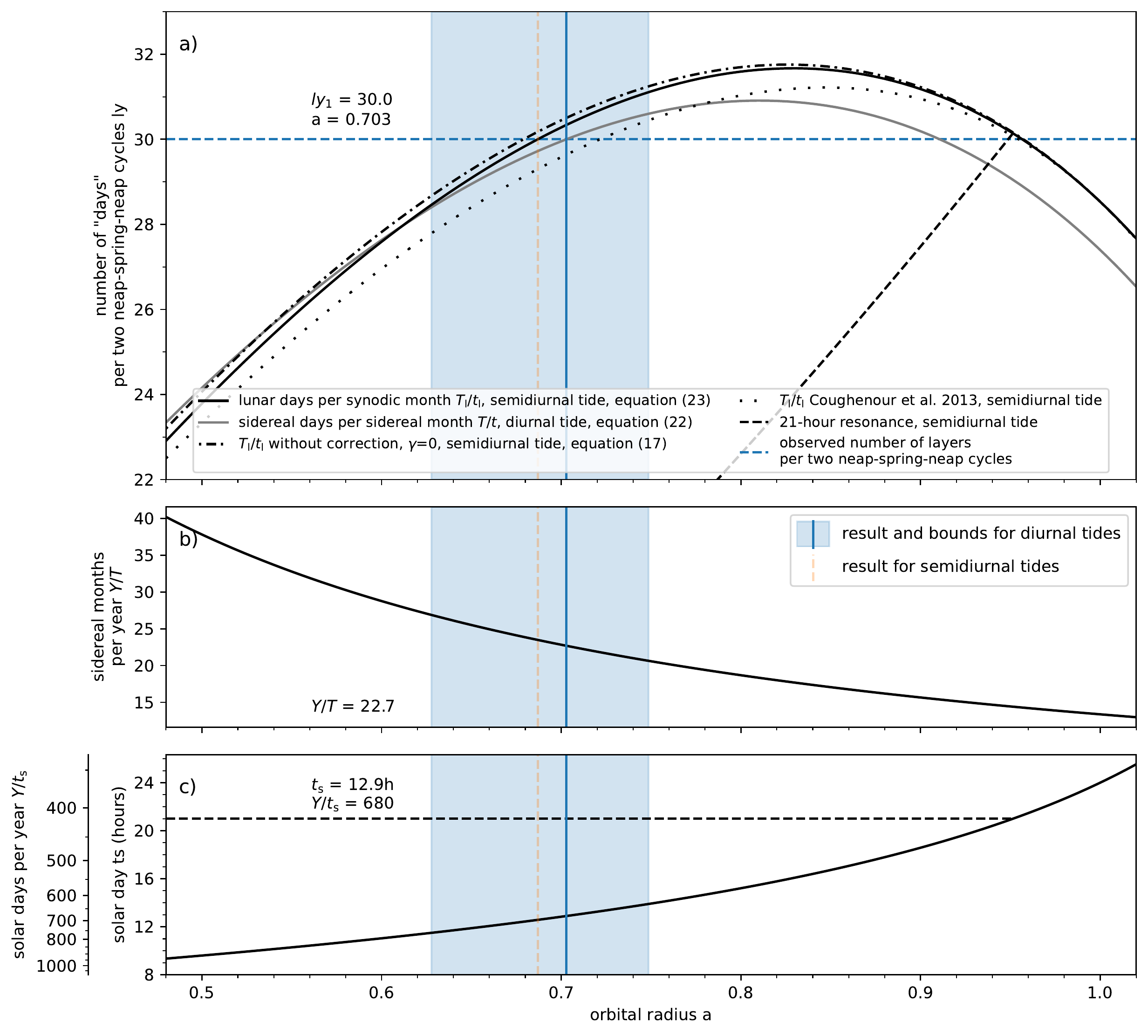}
\caption{
Relationship between different Earth-Moon dynamical parameters. The relationship between the observable \ly and orbital radius $a$ is different for diurnal and semidiurnal tidal deposits.
The observation from the tidal deposit in the Moodies Group $\lobs=30.0$ is marked by a horizontal dashed blue line in panel a.
Orbital parameters determined from this value assuming diurnal tides or semidiurnal tides are marked by vertical lines. Printed values and indicated bounds represent those of diurnal tides. 
}
\label{fig:orbit}
\end{figure}

For a given $\ly_1$ respective $\ly_2$, $a$ can be determined numerically by root finding.
The graphical representation of equation (\ref{eq:ly2a}) is displayed by a dot-and-dashed line in figure \ref{fig:orbit}a.

Equation~(\ref{eq:ly2a}) for semidiurnal tides is equivalent to equation (12) in \citet{Runcorn1979} after correcting a typo in \citet{Runcorn1979} by replacing $I/I_0/27.3$ with $27.3 I_0/I$, and after rolling back the substitutions $13.4\approx Y/T_0$, $27.3\approx T_0/t_0$ and $4.82\approx 1/\alpha_0$.

Differences in the dot-and-dashed line displayed in figure~1 of \citet{Runcorn1979} to our curve in figure~\ref{fig:orbit} can be explained by an outdated value \citet{Runcorn1979} used for Earth's moment of inertia, resulting in $\alpha_0=0.2075=1/4.82$.
\par 

After checking the units in equation~(13) of \citet{Coughenour2013}, we found that Earth's moment of inertia was not considered in the second term in the parenthesis of the left-hand side of their equation~(13).
We show the graph of the completed equation in figure~\ref{fig:orbit}a.
The curve is the same as shown in figure~5a of \citet{Coughenour2013}; clearly the moment of inertia missing in their equation was unintentional and the complete equation had been used in their calculations.

Another difference to our result for semidiurnal tides (equation~\ref{eq:ly2a}) consists in \citet{Coughenour2013} not integrating equation (\ref{eq:cons1}) but rather using simple differences. Therefore their equation is only a good approximation for orbital radii near today's value $a=1$ and for small orbital radii $a<0.1$.

\paragraph{Correction due to change in Earth's moment of inertia}

We add a small correction to equations (\ref{eq:ly1a}) and (\ref{eq:ly2a}) by considering a variation in the moment of inertia of Earth.
The younger, faster rotating Earth had a higher moment of inertia due to a higher bulge at the equator. Rotation therefore is slower than without this effect. Earth's polar moment of inertia can be expressed as \citep{Goldreich1966}:

\begin{align}
\Ie &= \mu\Iez\parent{1 + \gamma\parent{\frac{t_0^2}{t^2} - 1}} \text{ with} \label{eq:ie}\\
\gamma &= \frac{8\pi k_2^{\text f}r\ind e^5}{9G\Iez} = 0.00215 \text{ and } \mu = 1\
\end{align}

Here $r\ind e$ is Earth's radius, $G$ the gravitational constant and $k_2^{\text f}=0.93$ is Earth's fluid Love number around the rotation axis. $\gamma= 0.00215$ is a correction factor accounting for the change in Earth's polar moment of inertia due to a different rotation speed than today's. $\gamma=0$ corresponds to no change due to a different rotation speed of Earth. In this case equations~(\ref{eq:ly1a}) and (\ref{eq:ly2a}) can be used directly. The factor $\mu$ accounts for possible changes in moment of inertia independent from the rotational effect due to a different distribution of masses inside Earth.

Equation~(\ref{eq:ie}) can be inserted into equation~(\ref{eq:cons2}), leading to the cubic equation ($\gamma\neq 0$)
\begin{align}
\parent{\frac{t_0}{t}}^3 + p\parent{\frac{t_0}t} + q &= 0\text{ with} \label{eq:cube}\\
        p &= \frac{1 - \gamma}\gamma\nonumber\,,\\
        q &= -\frac{\X + \alpha_0}{\alpha_0\gamma\mu}\nonumber\,.

\end{align}

Equation (\ref{eq:cube}) has a single real root which can be calculated using Cardano's formula:

\newcommand{\C}{C\fof{a, \alpha_0, \beta_0, \gamma, \mu}}

\begin{equation}
\frac{t_0}t = \C =
\sqrt[3]{-\frac q2 + u} + \sqrt[3]{-\frac q2 - u} \text{ with } u = \sqrt{ \frac{q^2}4 + \frac{p^3}{27}} \label{eq:root}
\end{equation}

Finally, equations (\ref{eq:Kepler}), (\ref{eq:t1}), (\ref{eq:t2}) and (\ref{eq:root}) can be used to obtain

\begin{align}
\ly_1 &= \C \frac\exprT{t_0} \label{eq:ly1b}\\
\ly_2 &= \parent{\C \frac 1{t_0} - \frac 1\exprT} / \parent{\frac 1\exprT - \frac 1Y} \label{eq:ly2b}\,.
\end{align}

Again, for a given $\ly_1$ respective $\ly_2$, $a$ can be determined numerically by root finding.

\par
Furthermore, the rotation of Earth expressed as length of the solar day $\ts$ can be derived from the duration of a sidereal day $t$ and the duration of a year $Y$ by

\begin{equation}
\frac 1\ts = \frac 1t - \frac 1Y \,.\label{eq:ts}
\end{equation}

The evaluation of equations (\ref{eq:ly1b}) and (\ref{eq:ly2b}) as well as its simplified version (\ref{eq:ly2a}) are shown in figure \ref{fig:orbit}a.
Other orbital parameters related to the duration of the revolution of the Moon (equation~\ref{eq:Kepler}) and the rotation velocity of the Earth expressed as length of the solar day (equation~\ref{eq:t1} respective \ref{eq:t2} and equation~\ref{eq:ts}) are displayed as a function of orbital distance $a$ in figures \ref{fig:orbit}b and \ref{fig:orbit}c.

\section{Results and discussion}
\label{sec:final}

\begin{figure}
\centering
\includegraphics[width=0.95\textwidth]{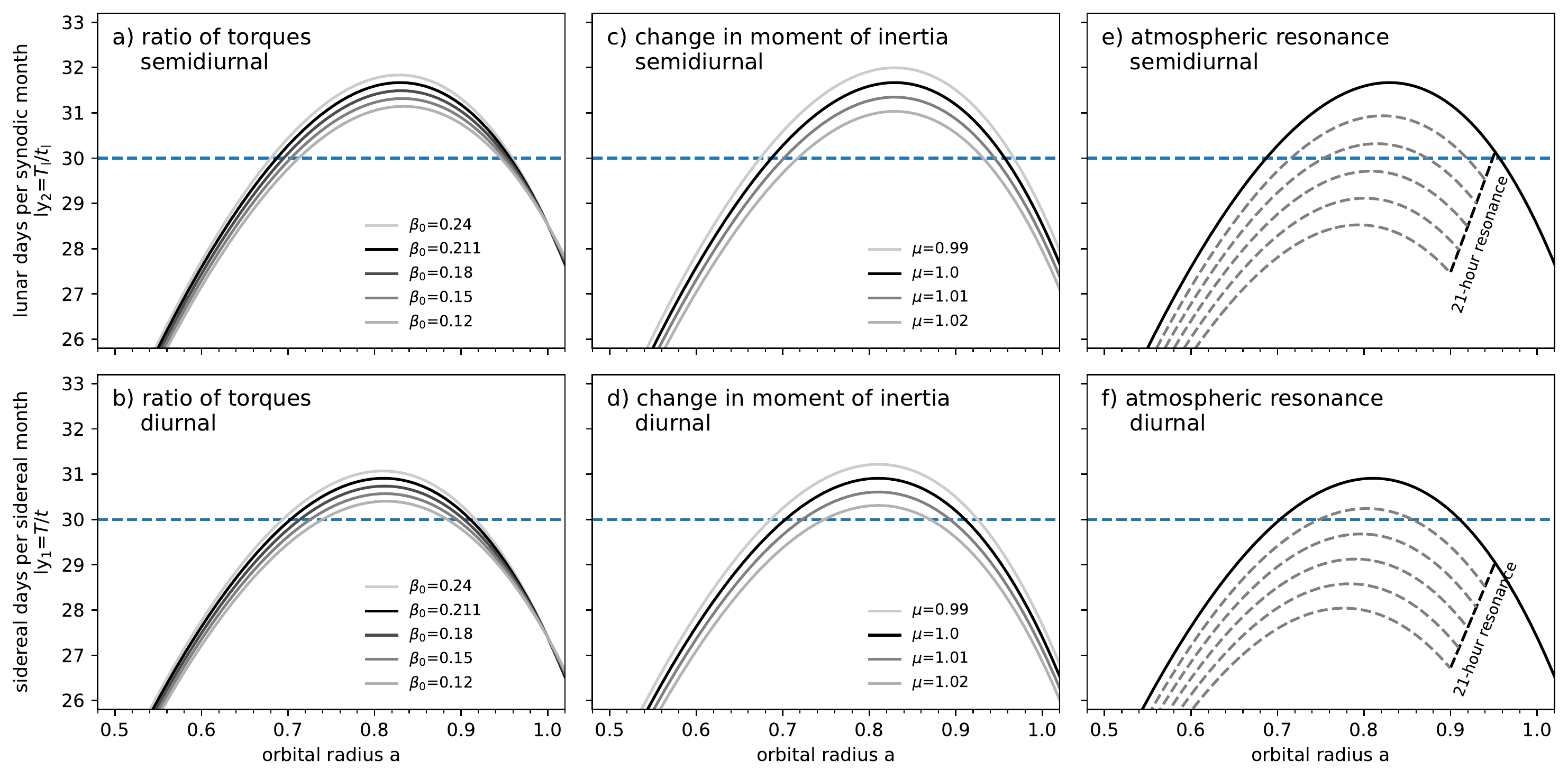}
\caption{
Top row: Relationship between the number of lunar days per synodic month ($\ly_2=\Tl/\tl$) and orbital radius $a$ for semidiurnal tidal deposits.\newline
Bottom row: Relationship between the number of sidereal days per sidereal month ($\ly_1=T/t$) and orbital radius $a$ for diurnal tidal deposits. The original graphs are displayed by black lines. The observation $\ly\ind{obs}=30.0$ is indicated by the dashed blue line.\newline
a) Relationship between $\ly_2$ and $a$ for different ratios of today's solar torque to today's lunar torque ($\beta_0$), with values ranging from 0.12 to 0.24. The ratio of solar torque to lunar torque is assumed to change in time as a function of orbital radius $a$, as described by equation~(\ref{eq:constLs}).\newline
b) Same plot as in a), but the y axis displays $\ly_1$. The orbital radius derived from our observation varies between 69\% and 74\%, depending on the value of $\beta_0$.\newline
c) Relationship between $\ly_2$ and $a$ for a change in Earth's moment of inertia adding to the rotational effect. We show curves for an additional change in moment of inertia between -1\% and 2\% ($0.99\leq \gamma\leq1.02$).\newline
d) Same plot as in a), but the y axis displays $\ly_1$. For example, the orbital radius derived from our observation for $\gamma=1.01$ is 72\% of today's value.\newline
e) Relationship between $\ly_2$ and $a$ for different durations of the period in which Earth was trapped in the 21-hour atmospheric resonance. The original curve without resonance is shown by a continuous black line. The black dashed line represents the orbits in which Earth was trapped in the 21-hour resonance until reaching an orbital radius of around 95\% of the present value. Gray dashed lines represent the time before entering the resonance at 90\%, 91\%, 92\%, 93\% and 94\% of today's orbital radius.\newline
f) Same plot as in c), but the y axis displays $\ly_1$. Introduction of the 21-hour resonance generally results in higher estimates of orbital radius for our observation.
}
\label{fig:orbit_params}

\end{figure}

The results of this study are based on two assumptions: 
Firstly, the amount of angular momentum leaked from the Earth-Moon system proportional to the ratio of tide-raising torques from Sun and Moon can be expressed by equation~(\ref{eq:constLs}).
The proportionality to $a^6$ is founded in the proportionality of tidal amplitudes to inter-body distance cubed, considering the second-order nature of tidal dissipation and a nearly constant Earth-Sun distance \citep[e.g.][]{Goldreich1966b}. Equation~(\ref{eq:constLs}) also requires lag angles of lunar and solar contribution to the tidal torque to be similar and that the torques are proportional to the tidal forces \citep{Brosche1990}. These assumptions are strictly valid only for equilibrium tides. For example, estimates of today's ratio between solar and lunar tidal torques range from 0.13 to 0.19 \citep[p.~457]{Zharkov1996}, compared to the theoretical value of $\beta_0=0.211$ used in equation~(\ref{eq:constLs}). The discrepancy between these values arises from the different responses of the ocean to today's $M_2$ and $S_2$ tide due to their slightly different frequencies. Because ancient oceans with different geometries had other eigenperiods and the Earth's rotation and the frequencies of the tides were different, equation~(\ref{eq:constLs}) may nevertheless be valid in average. Assuming different values of the solar-to-lunar torque ratio will change the relationship between $\ly$ and $a$ (figure~\ref{fig:orbit_params}a-b). 
\citet{Deubner1990} estimated an Earth-Moon distance from the Elatina Formation, 620 million years old, using equation~(\ref{eq:cons2}) and $\beta_0=0.211$ in equation~(\ref{eq:constLs}), consistent with both estimates using Kepler's law and the length of the lunar nodal cycle, indicating the validity of the equations for the Elatina tidal record.

A variation in tilt of Earth's rotation axis, which is currently 23.4°, is also neglected. This assumption was violated in the early solar system, but precise calculations of \citet{Goldreich1966} and \citet{Rubincam2016} suggest obliquity variations only down to around 20° for $a=0.6$.
The small change in obliquity for $a\ge 0.6$ suggests that it can be ignored in our study. Additionally, we numerically verified our $ly_1=T/t$ and $ly_2=T_l/t_l$ versus $a$ curves with the calculations of \citet{Goldreich1966} by digitizing the $\ts-a$ curve in figure 6 of \citet{Goldreich1966} and applying the equations (\ref{eq:Kepler}), (\ref{eq:t1}), (\ref{eq:t2}), (\ref{eq:ts}).

\par

The second assumption is about the tidal character of the deposited layers. \citet{Eriksson2000} and \citet{Heubeck2022} both interpret the investigated rock outcrop of Moodies Group sandstone, exposing a sequence of foreset layers from a large subaqueous dune, to have formed due to rhythmically varying tidal currents in a subtidal environment. This is supported by measured bidirectional, opposing paleocurrent directions in under- and overlying strata which indicate reversing currents, the presence of mud chips indicating erosion of dried-out, suspension-settled mud nearby, the composition of the medium-grained, well sorted, quartz-rich sandstone free of gravel, and the overall setting between underlying deposits of a sandy coastal plain and overlying strata of marine affinity.

It is necessary to consider the different imprint of diurnal and semidiurnal dominated tides on the duration of the neap-spring-neap cycle and the subsequent interpretation of the observed number of layers per two neap-spring-neap cycles, because of its possible and plausible relation to diurnal tides.\par

\begin{table}
  \caption{Parameters describing the lunar orbit and Earth's rotation rate 3.2 billion years ago derived from the primary observation $\lobs{=}30.0$ from the Moodies tidal record, using various assumptions.
  For the columns assuming diurnal dominating tides we use $\ly_1{=}\lobs$ and equation~(\ref{eq:ly1b}), for semidiurnal dominating tides $\ly_2{=}\lobs$ and equation~(\ref{eq:ly2b}), for the 21-hour atmospheric resonance we use $\ly_2{=}\lobs/2$.
  Lower and upper bounds (in parentheses) are shown for diurnal dominating tides and default assumptions. For comparison, today's values are shown to the right.}
  \label{tab:res}

   \begin{small}\makebox[\textwidth][c]{\begin{tabular}{llllllll}
    \toprule
    \multirow{3}{*}{assuming} & \multicolumn{4}{l}{diurnal tides} & \multicolumn{2}{l}{semidiurnal tides} & today\\

    & \multirow{2}{*}{default (bounds)} & \multirow{2}{*}{$\gamma{=}0^a$} &
    $\mu{=}$   & high & \multirow{2}{*}{default} &\SI{21}{h} & \\
    &              & & $1.01^b$   &  orbit$^c$ & &resonance &\\
    \midrule
    $\ly_1=T/t$ &                   30.0 (28.4, 30.6)    &  30.0 &  30.0 &  30.0 &  29.7 &  15.4 & 27.4\\
    $\ly_2=\Tl/t\ind l$ &             30.3 (28.5, 31.1)    &  30.3 &  30.4 &  31.0 &  30.0 &  15.0 & 28.5\\
    Moon-Earth distance $a$ &            0.703 (0.628, 0.749) &  0.693 & 0.723 & 0.911 & 0.687 & 0.624 & 1.000\\
    Sid.\ months per year $Y/T$ &         22.7 (26.9, 20.6)    &  23.3 &  21.8 &  15.4 &  23.5 &  27.1 & 13.4\\
    Sol.\ days p.\ syn.\ month $\Tl/\ts$ &31.3 (29.5, 32.1)    &  31.3 &  31.4 &  32.0 &  31.0 &  16.0 & 29.5\\
    Solar day $\ts$ (hours) &             12.9 (11.5, 13.9)    &  12.6 &  13.4 &  19.0 &  12.6 &  21.0 & 24.0\\ 
    Solar days per year $Y/\ts$ &         680  (762, 631)      &  694 &   652 &   460 &   697 &   417 & 365\\
    \bottomrule    
  \end{tabular} 
  }\end{small} 
  
    {\footnotesize
      $^a$Using equation (\ref{eq:ly1a}) instead of equation (\ref{eq:ly2b}), ignoring the correction for the rotational effect in $\Ie$\newline
      $^b$Assuming a 1\% higher moment of inertia $\Ie$ adding to the rotational effect\newline
      $^c$Using the second root in equation (\ref{eq:ly2b}) with $a>0.81$\par
      }

\end{table}

In section~\ref{sec:specs} we estimated the number of layers per two neap-spring-neap cycles as $\lobs=30.0$ (with lower and upper bounds of 28.4 and 30.6, respectively), taking missing layers for parts of the data set into account. The true value might be higher due to potential hiatuses in the short sequence of around 100 measurements. As discussed in section~\ref{sec:tides}, $\lobs=\ly$ represents the number of lunar or sidereal days per two neap-spring-neap cycles.

\begin{figure}
\centering
\includegraphics[width=0.7\textwidth]{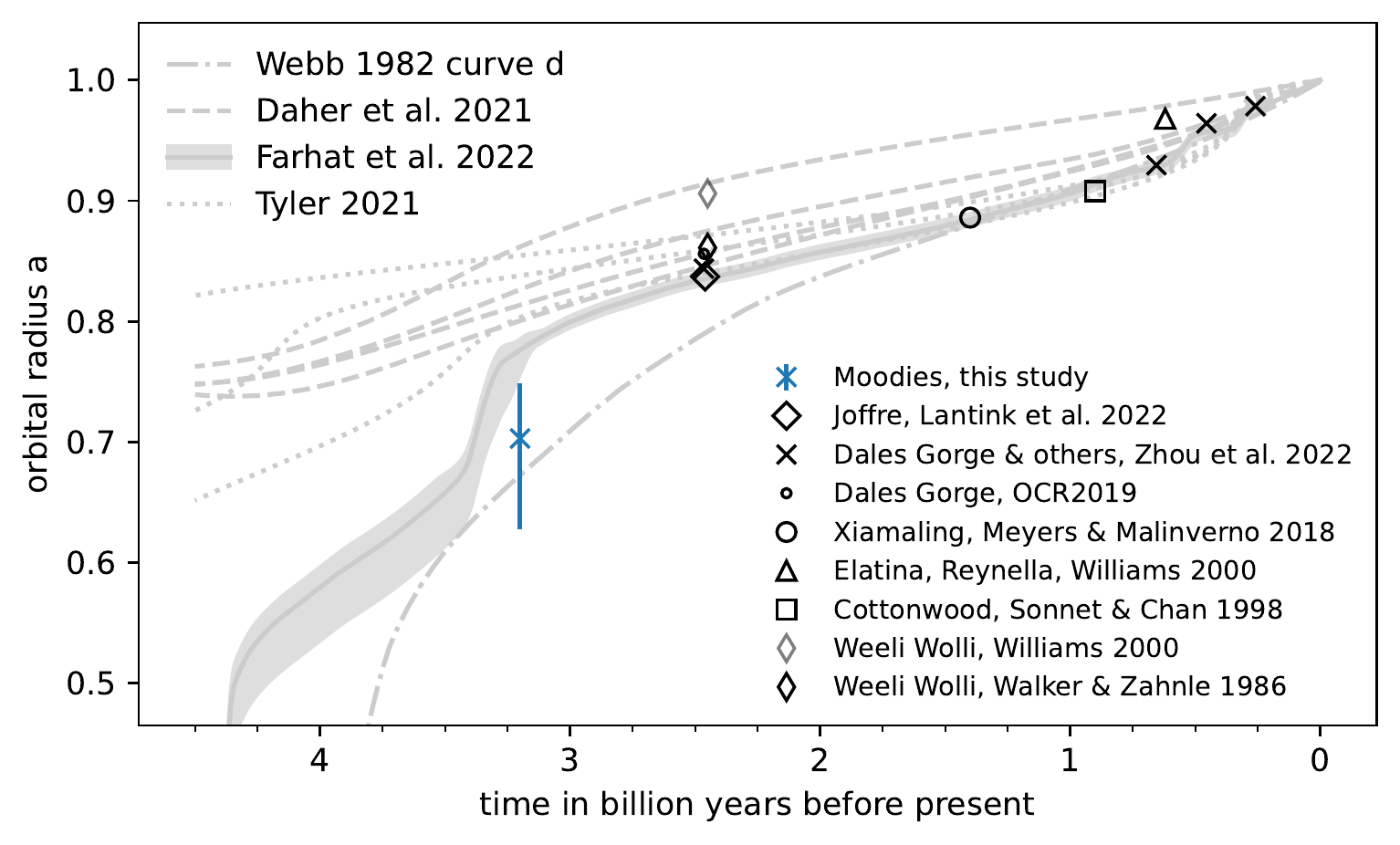}
\caption{Evolution of orbital radius $a$ relative to today's value. Displayed are different data points together with our estimate assuming dominant diurnal tides. The value reported in \citet{OliveiraCarvalhoRodrigues2019} is labeled with OCR2019. Lines represent the model of \citet{Webb1982} taking into account power dissipation in both the ocean and solid Earth (curve d), the ocean models of \citet{Daher2021} with modern ocean basin geometry and with reconstructed ocean basin geometries 55, 116, and 252 million years ago, the model of \citet[figure~4b]{Tyler2021} for dissipation time scales of 30, 40 and 50 and the model of \citet{Farhat2022} with 95\% confidence intervals for the determination of the parameter.
}
\label{fig:orbit_time}
\nocite{Williams2000, Sonett1998, Meyers2018, Walker1986, Zhou2022, Lantink2022}
\end{figure}

Assuming dominant diurnal tides, we estimate an Earth-Moon distance of around 70\% (with lower and upper bounds of 63\% and 75\%, respectively) of today's value 3.2 billion years ago, and a year with around 680 solar days of ca.\ 13 hours duration each (figure~\ref{fig:orbit}). Figure~\ref{fig:orbit_time} displays this estimate in the context of other estimates derived from younger strata. Table~\ref{tab:res} shows detailed results for different assumptions. Ignoring the change of Earth's equatorial bulge due to its faster rotation ($\gamma=0$) yields a slightly lower estimate of the Earth-Moon distance of 69\% (table~\ref{tab:res}). Allowing for an additional higher moment of inertia of Earth independent from the rotational effect results in higher estimates. E.g., for an additional increase in the moment of inertia of 1\% ($\mu=1.01$), the Earth-Moon distance would increase to around 72\% of today's value (figure~\ref{fig:orbit_params}d, table~\ref{tab:res}).

In contrast, assuming dominant semidiurnal tides, we estimate an Earth-Moon distance of around  69\% of today’s value (table~\ref{tab:res}); this is identical within the limits of uncertainty to the value above derived from the assumption of diurnal tides because the relation $ly-a$ for diurnal and semidiurnal tides at this distance $a$ is similar.

Using the completed equation of \citet{Coughenour2013} for semidiurnal tides yields similar results but introduces a relative error larger than 5\% in the estimate of $a$. Because the slope of the $\ly-a$ relations in equations (\ref{eq:ly1b}) and (\ref{eq:ly2b}) is relatively low, differences in the estimate of $\ly$ result in large changes in the estimate of $a$. Furthermore, the non-uniqueness of the $\ly-a$ relation allows for a second solution resulting in a higher orbit at 91\% of today's distance assuming diurnal tides (table~\ref{tab:res}). 

This result is, however, implausible because the similar distance $a$ derived from the much younger Elatina Formation (620~million years) would have required a pause in the evolution of the lunar orbit for more than two billion years.
\par

The reconstructed Earth-Moon distance 3.2 billion years ago would be moderately altered to between 69\% and 74\% of today’s value assuming diurnal tides when employing values between 0.12 and 0.24 for the ratio between today’s solar and lunar tide-raising torque (figure~\ref{fig:orbit_params}b).

\citet{Zahnle1987} suggested that the length of the solar day remained fixed at 21 hours for a long period in the Precambrian due to a resonance between solar-induced atmospheric tides and the Earth's rotational period, counteracting the deceleration of Earth's rotation due to tidal friction.
For the time period in which Earth would have been trapped in the resonance, Earth's angular momentum is constant, $\diff\Le=0$, contrary to equation~(\ref{eq:consttotal}). The $\ly-a$ relation can then simply be derived from equations (\ref{eq:t1}), (\ref{eq:t2}) and (\ref{eq:ts}) and is shown in figure~\ref{fig:orbit} assuming the Earth was trapped in the resonance until Moon reached an orbit of $a=0.95$.
A good fit to this curve can only be obtained for a ratio $\lobs/\ly=2$ (assuming semidiurnal tides) at a low $a$ value of 62\% (table~\ref{tab:res}). It is thus likely that the 21-hour resonance either occurred for a short duration or not at all. The time period before Earth was possibly trapped in the 21-hour resonance can again be constrained by our equations if a small change is incorporated. The integration in equation~(\ref{eq:integral}) has to be calculated from the orbit $a_1$ in which Earth entered the resonance to $a$, not from $a_0{=}1$ to $a$. Therefore, $\X$ can be determined by substituting $a_0$ by $a_1$ in equation~(\ref{eq:X}). Correspondingly, $t_0$ has to be substituted by $t_1=1/\parent{1/\SI{21}h + 1/Y}=\SI{20.95}h$.

Figure~\ref{fig:orbit_params}e-f displays the $ly-a$ curve for different durations of the 21-hour resonance for semidiurnal and diurnal tides.

The curves correspond to orbits between 90\% and 94\% of today's Earth-Moon distance for which the Earth entered the 21-hour atmospheric resonance. Assuming diurnal tides and the occurrence of the 21-hour resonance of unknown duration and taking into account the non-uniqueness of the $\ly-a$ relationship, any value between 70\% and 91\% for the orbital radius 3.2 billion years ago would be consistent with the observation $\lobs=30.0$ (figure~\ref{fig:orbit_params}d). Only a short duration of the 21-hour resonance would be compatible with our observation based on diurnal tides, because otherwise the maxima of the $ly-a$ relationship would be lower than 30. The corresponding orbital range of $0.935<a<0.95$ corresponds to a maximum duration of 200 million years for the 21-hour resonance using the model of \citet{Webb1982}.

\par

The estimate of this study and the data points of \citet{Sonett1998} and \citet{Meyers2018} fit well the model of \citet{Webb1982} which takes into account power dissipation in the ocean and solid Earth (figure~\ref{fig:orbit_time}). The back-projection of \citeauthor{Webb1982}'s model leads to an age of the Moon of 3.9 billion years which is somewhat too low.

The ocean model of \citet{Daher2021} using ocean configurations of the present day and 55 million, 116 million and 252 million years ago fits less well because it predicts a high value of 75\% of today’s Earth-Moon distance at the time of the formation of the Moon 4.5 billion years ago.

Results from the Weeli Wolli Formation \citep{Walker1986, Williams2000} fit the model of \citet{Daher2021}, but the tidal nature of the Weeli Wolli strata is debated \citep[e.g.][]{Trendall1973}.

The orbital model of \citet{Tyler2021}, based on effective ocean depth and dissipation timescale, was constrained by observations of tidal laminae including the Weeli Wolli data point. Our result would be consistent with \citeauthor{Tyler2021}'s model using a dissipation time scale shorter than 30 or by adding additional dissipation due to solid Earth tides.

The orbital model of \citet{Farhat2022}, only constrained by today's recession rate of the Moon and the time of its origin, fits several data points in figure~\ref{fig:orbit_time} well. The uncertainty shown reflects the 95\% confidence intervals for the determination of the parameter; the model's inherent error might be larger.
A defining feature of the model of \citet{Farhat2022} are different tidal resonances -- one of these tidal resonances leads to a drastic increase in Earth-Moon distance of more than 10\% in less than 200 million years around 3.3 billion years ago.
Our estimate for Archean Earth-Moon distance, being smaller than in the model of \citet{Farhat2022}, might indicate that this tidal resonance took place later than predicted.

\section{Conclusions}

We present an updated analysis of the sole geological data point constraining Archean Earth-Moon dynamics, a rock outcrop of tidal strata from the 3.22 billion years old Moodies Group of South Africa.
By taking into account the possibility of missing layers, we obtain a primary value of 30.0 layers per two neap-spring-neap cycles.
The implications for the lunar orbit depend on whether the deposition was dominated by diurnal or semidiurnal tides. The derivation of the relationship between the number of layers per two neap-spring-neap cycles and the lunar orbit also demonstrates the significance of leaking angular momentum from the Earth-Moon system.
A small correction for the change in Earth's moment of inertia due to its changing rotation speed is introduced.
We compare our results to earlier work \citep{Coughenour2013} and explain the differences between the two studies.
Assuming dominating diurnal tides, the Earth-Moon distance 3.2 billion years ago is estimated at around 70\% of today's value.
Different ratios of solar and lunar torques, small changes in Earth's moment of inertia independent of the rotational effect, and the assumption of a deposition due to semidiurnal dominating tides (as opposed to diurnal dominating tides) alter this value moderately.
A duration of a postulated 21-hour atmospheric resonance shorter than 200 million years would be consistent with our observation, but it would significantly alter our estimate of the Archean Earth-Moon distance.
Analysis of tidal deposits from recent scientific drilling in the Moodies Group may further constrain the reported estimate.

\paragraph{Open research}
\begin{footnotesize}
Processing and plotting was performed with the libraries SciPy, NumPy and matplotlib \citep{SciPy, NumPy, Hunter2007}. The synthetic tidal signals in figure~\ref{fig:tides} were generated with ETERNA predict via its Python wrapper pygtide \citep{Wenzel1996, Rau2022}.
This article can be reproduced with the data and source code provided at \url{https://github.com/trichter/archean_moon_orbit} \citep{Eulenfeld2022_moon_sourcecode}. 
The model of \citet{Daher2021} can be downloaded from \citet{Arbic2021}.
\end{footnotesize}

\paragraph{Acknowledgments}
\begin{footnotesize}
We thank Richard Ray for sharing the model of \citet{Webb1982}, Jacques Laskar for providing data of the model of \citet{Farhat2022} and Robert Tyler for providing data of his model.
We thank Stephen Mojzsis, Christopher Coughenour and two anonymous reviewers for comments on an early version of the manuscript.
\end{footnotesize}

\begin{footnotesize}
\setlength{\bibsep}{1.5ex}
\catcode`\^^M=5

\end{footnotesize}

\end{document}